\documentstyle[psfig]{aipproc}
\def\M{{\cal M}}
\def\be{\begin{equation}}
\def\ee{\end{equation}}
\def\bea{\begin{eqnarray}}
\def\eea{\end{eqnarray}}

\begin{document}
\title{ LISA observations of massive black hole binaries
    using post-Newtonian waveforms}
\author{Alicia M. Sintes and Alberto Vecchio}
\address{Max-Planck-Institut f\"ur Gravitationsphysik,
 Am M\"uhlenberg 1,  D-14476 Golm, Germany}
\maketitle
\begin{abstract}
We consider LISA observations of in-spiral signals emitted by
massive black hole binary systems in circular orbit and
with negligible spins. We study the accuracy with which
the source parameters can be extracted from the data stream.
We show that the use of waveforms retaining post-Newtonian 
corrections not only to the phase  but also the amplitude
can drastically improve the estimation of some parameters.
\end{abstract}
%%%%%%%%%%%%%%%%%%%%%%%%%%%%%%%%%%%%%%%
The strongest sources of gravitational  waves for LISA \cite{lisa,V}
 are likely  to be
binary systems of massive black holes ($10^4$-$10^7$ M$_{\odot}$).
%as
%they  would be detectable, in the final year of in-spiral,
%at a  signal-to-noise ratio SNR $\sim$ $10^2$-$10^4$ for sources at 
%cosmological distances.
%Using the post-Newtonian (PN) approximation scheme to general relativity,
%the waveform from in-spiral compact binaries can be written schematically
%through 2PN order as \cite{Bla}:
%\bea
%h(t)=&{\rm Re}& \left[ (h_1^{0.5}+h_1^{1.5}+h_1^2) \  e^{i \Phi} 
%+ (h_2^0+h_2^1+h_2^{1.5}+h_2^2) \  e^{i 2\Phi} \right. \nonumber\\
%& & + \left. (h_3^{0.5}+h_3^{1.5}+h_3^2) \  e^{i 3\Phi} 
% +(h_4^1+h_4^2) \  e^{i 4\Phi}
%+ h_5^{1.5}\  e^{i 5\Phi} 
%  + h_6^2\  e^{i 6\Phi} \right] \ , 
%\eea
%where the superindex indicates the term's PN order, the 
%subindex labels the different harmonics, and $\Phi$ has the PN expansion
%$\Phi (t)=\Phi^0 + \Phi^1+ \Phi^{1.5}+ \Phi^2$.
%%%%%%%%%%%%%%%%%%%%%%%%%%%
So far, parameter estimation of in-spiral signals for LISA 
has been
investigated
within the so-called restricted post-Newtonian approximation
\cite{Curt,VC}: PN corrections
are taken into account in the phase of the waveform, whereas the amplitude
is retained at the lowest Newtonian order. 
%Thus, one discards all 
%multipole components except the quadrupole one.
Here, we investigate the implications for parameter estimation of 
the introduction of PN corrections also to the amplitude. Going to higher
PN order in the amplitude implies the use of several multipole components
and not just the quadrupole one.
%We consider a binary source of masses $m_1$ and $m_2$, that we assume 
%in circular orbit   and with negligible spins, i.e., whose
%contribution is accounted for only in the GW phase.
The gravitational radiation is then described by eleven 
independent parameters associated with distance, masses,
 spins, position and orientation of the source in the sky,
and  instant and phase of the final collapse.
%%%%%%%%%%%
%%We consider the case of LISA observations of massive black hole binaries
%%for a time of observation  corresponding to the final year of in-spiral.
We compute the results regarding the expected errors  associated  with 
measurement of the parameters that characterize the source using two different
waveform approximations \cite{Bla}:
(a) The standard restricted 2 PN approximation:
$h(t)={\rm Re}\left[ \, h_2^0 \  e^{i 2\Phi}  \right]$;
(b) A signal where we consider radiation emitted not only at twice the 
orbital frequency $f_{orb}$ but also at  $f_{orb}$  and $3f_{orb}$;
in this case we retain only the 0.5 PN correction to the amplitude,
the signal reads:
$h(t)={\rm Re} \left[ \, h_1^{0.5} \  e^{i \Phi} \ + \ h_2^0 \  e^{i 2\Phi}
\ + \ h_3^{0.5} \  e^{i 3\Phi}\right]$.
The results depend strongly on the actual values of the source
parameters, in particular  location and orientation of the source.
 Therefore, we perform a Monte-Carlo simulation, 
keeping the source distance and masses fixed and varying  randomly
$\hat {{\bf N}}$ and $\hat {{\bf L}}$, and we construct histograms to show 
the distributions of the different parameter measurement errors.
\begin{figure}[h!]
\centerline{\vbox{ 
\psfig{figure=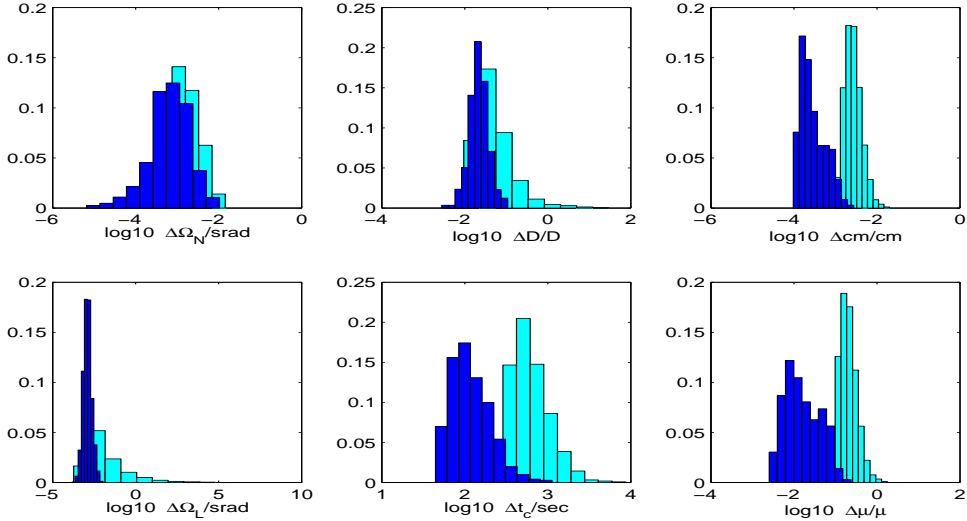,height=7cm,width=13cm} 
}} 
\vspace{5pt}
\caption{
Probability distribution of the angular resolution, orientation,
distance measurement errors,  timing accuracy and
mass measurements errors for LISA observations of the final year of
massive black hole in-spiral binaries 
($m_1= 10^7 \, M_{\odot}$, $m_2= 10^6 \, M_{\odot}$, $z=1$).
The light and dark colors refer to waveform (a) and (b) respectively
(see text).
%In light color,
% the waveform is: $h={\rm Re} \left[ h_2^0 \  e^{i 2\Phi}\right] $.
%In dark, the considered waveform model is:
%$h={\rm Re} \left[h_1^{0.5} \  e^{i \Phi} \ + \ h_2^0 \  e^{i 2\Phi}
%\ + \ h_3^{0.5} \  e^{i 3\Phi}\right] $.
The histograms show the error distribution for 1000 random values of 
$\hat {{\bf N}}$ and $\hat {{\bf L}}$.
% and are given for the case of
%measurements with only one interferometer.
} 
%\vspace*{10pt}
\end{figure} 
We checked that  the angular resolution
$\Omega_N$ is basically unaffected by adding amplitude corrections.
Even though in the waveform (b) there is more information about
position and orientation of the source, this turned out to be too
weak to provide an improvement  in the angular resolution.
For the distance $\Delta D/D$,  we have the same values on average but 
smaller dispersion of the errors around the mean.
 On the contrary, the determination of physical parameters such as
 time of coalescence $t_c$, chirp mass $\M$ 
 (referred as cm in the figure), and
 reduced mass $\mu$ is strongly improved by a factor of 10 ore more:
 Within model (a) information about $t_c$, $\M$ and  $\mu$ are 
conveyed only by the GW phase  $2\Phi$; in model (b)
  the evolution of all three phases and  amplitudes 
 is controlled in a different way by
 those physical parameters  providing therefore more information.

 We conclude that according to our results, 
 more accurate (i.e., beyond the standard restricted PN approximation)
 in-spiral waveforms,
 as those
 including the radiation  emitted at different  multiples of the 
 orbital frequency, do play an important role for LISA 
 as GW observatory; in fact 
  they  provide extra information that  improve the 
 measurements  of the physical source parameters.

\end{document}